\input{aipcheck}

\documentclass[
    ,final            
  ]
  {aipproc}

\layoutstyle{6x9}
\usepackage{axodraw}
\begin{document}

\title{Dark Left-Right Model: CDMS, LHC, etc. 
\\}


\author{Ernest Ma}{
  address={Physics and Astronomy Department, University of California, 
Riverside, California 92521, USA}
}

\begin{abstract}
The Standard Model of particle interactions is extended to include 
fermion doublets $(n,e)_R$ transforming under the gauge group $SU(2)_R$ 
such that $n$ is a Dirac scotino (dark-matter fermion), with odd $R$ parity. 
Based on recent CDMS data, it is shown how this new dark left-right model 
(DLRM2) favors a $Z'$ gauge boson at around 1 or 2 TeV and be observable 
at the LHC.  The new $W_R^\pm$ gauge bosons may also contribute significantly 
to lepton-flavor-changing processes such as $\mu \to e \gamma$ and $\mu-e$ 
conversion in a nucleus or muonic atom.
\end{abstract}

\maketitle

\section{Left-right Extension of Standard Model}
If the Standard Model (SM) of particle interactions is extended to 
accommodate $SU(3)_C \times SU(2)_L \times SU(2)_R \times U(1)_X$, then 
the conventional assignment of
\begin{eqnarray}
(\nu,l)_L \sim (1,2,1,-1/2), &~& (\nu,l)_R \sim (1,1,2,-1/2), \\ 
(u,d)_L \sim (3,2,1,1/6), &~& (u,d)_R \sim (3,1,2,1/6),
\end{eqnarray}
implies the well-known result that $X = (B-L)/2$ and $Y = T_{3R} + (B-L)/2$. 
There must then be Higgs bidoublets
\begin{equation}
\Phi = \pmatrix{\phi_1^0 & \phi_2^+ \cr \phi_1^- & \phi_2^0}, ~~~ 
\tilde{\Phi} = \pmatrix{\bar{\phi}_2^0 & -\phi_1^+ \cr -\phi_2^- & 
\bar{\phi}_1^0},
\end{equation}
both transforming as $(1,2,2,0)$, yielding lepton Dirac mass terms 
\begin{equation}
m_l = f_l \langle \phi_2^0 \rangle + f'_l \langle \bar{\phi}_1^0 \rangle, ~~~
m_\nu = f_l \langle \phi_1^0 \rangle + f'_l \langle \bar{\phi}_2^0 \rangle,
\end{equation}
and similarly in the quark sector.  This results in the appearance of 
phenomenologically undesirable tree-level flavor-changing neutral currents 
from Higgs exchange, as well as inevitable $W_L - W_R$ mixing.  If 
supersymmetry is imposed, then $\tilde{\Phi}$ can be eliminated, but then 
$({M}_\nu)_{ij} \propto ({M}_l)_{ij}$ as well as 
$({M}_u)_{ij} \propto ({M}_d)_{ij}$, contrary to what is observed. 
Hence the prevalent thinking is that $SU(2)_R \times U(1)_{B-L}$ is actually 
broken down to $U(1)_Y$ at a very high scale from an $SU(2)_R$ Higgs triplet 
$(\Delta_R^{++},\Delta_R^+,\Delta_R^0) \sim (1,1,3,1)$ which provides $\nu_R$ at 
the same time with a large Majorana mass from $\langle \Delta_R^0 \rangle$. 

The canonical seesaw mechanism for neutrino mass is thus implemented and 
everyone should be happy. But wait, no remnant of the $SU(2)_R$ gauge 
symmetry is detectable at the TeV scale and we will not know if $\nu_R$ 
really exists.  Is there a natural way to lower the $SU(2)_R \times 
U(1)_{B-L}$ breaking scale?

The answer was already provided 23 years ago~\cite{m87} in the context of 
the superstring-inspired supersymmetric $E_6$ model.  The fundamental 
\underline{27} fermion representation here is decomposed under 
$[(SO(10),SU(5)]$ as
\begin{equation}
\underline{27} = (16,10) + (16,5^*) + (16,1) + (10,5) + (10,5^*) + (1,1).
\end{equation}
Under its maximum subgroup $SU(3)_C \times SU(3)_L \times SU(3)_R$, the 
\underline{27} is organized instead as $(3,3^*,1)+(1,3,3^*)+(3^*,1,3)$, i.e.
\begin{equation}
\pmatrix{d & u & h \cr d & u & h \cr d & u & h} + \pmatrix{N & E^c & \nu \cr 
E & N^c & e \cr \nu^c & e^c & n^c} + \pmatrix{d^c & d^c & d^c \cr u^c & u^c 
& u^c \cr h^c & h^c & h^c}.
\end{equation}
It was realized~\cite{m87} in 1987 that there are actually two left-right 
options: (A) Let $E_6$ break down to the fermion content of the conventional 
$SO(10)$, given by $(16,10)+(16,5^*)+(16,1)$, which is the usual 
left-right model which everybody knows. (B) Let $E_6$ break down to the 
fermion content given by $(16,10)+(10,5^*)+(1,1)$ instead, thereby switching 
the first and third rows of $(3^*,1,3)$ and the first and third columns of 
$(1,3,3^*)$. Thus $(\nu,e)_R$ becomes $(n,e)_R$ and $n_R$ is not the mass 
partner of $\nu_L$. This is referred to by the Particle Data Group as the 
Alternative Left-Right Model (ALRM).  Here the usual left-handed lepton 
doublet is part of a bidoublet:
\begin{equation}
\pmatrix{\nu & E^c \cr e & N^c}_L \sim (1,2,2,0).
\end{equation}
In this supersymmetric model, $\nu_L$ is still the Dirac mass partner of 
$\nu_R$ and gets a seesaw mass, whereas~\cite{m00} $n_R$ (which couples to 
$e_R$ through $W_R$) mixes with the usual neutralinos, the lightest of which 
is a dark-matter candidate.

\section{Dark left-right model}
Two simpler nonsupersymmetric versions of the ALRM with $n_R$ as dark matter 
have recently been proposed~\cite{klm09,klm10}.  We call them Dark Left-Right 
Models (DLRM and DLRM2).  We impose a global U(1) symmetry $S$, so that 
under $SU(3)_C \times SU(2)_L \times SU(2)_R \times U(1)_X \times S$, where 
$Q = T_{3L} + T_{3R} + X$, a generalized lepton number is conserved, such 
that $L = S - T_{3R}$ in DLRM and $L = S + T_{3R}$ in DLRM2.  The resulting 
dark-matter fermion $n_R$ has $L=0$ (Majorana) in DLRM and $L=2$ (Dirac) 
in DLRM2.  This talk is on DLRM2, with particle content~\cite{klm10} 
under $SU(3)_C \times SU(2)_L \times SU(2)_R \times U(1)_X \times S$ given 
below:
\begin{eqnarray}
&& \psi_L = (\nu,e)_L \sim (1,2,1,-1/2;1), ~~~ \nu_R \sim (1,1,1,0;1), \\  
&& \psi_R = (n,e)_R \sim (1,1,2,-1/2;3/2), ~~~ n_L \sim (1,1,1,0;2), \\
&& Q_L = (u,d)_L \sim (3,2,1,1/6;0), ~~~ d_R \sim (3,1,1,-1/3;0), \\ 
&& Q_R = (u,h)_R \sim (3,1,2,1/6;-1/2), ~~~ h_L \sim (3,1,1,-1/3;-1), \\ 
&& \Phi \sim (1,2,2,0;-1/2), ~~~ \tilde{\Phi} \sim (1,2,2,0;1/2), \\ 
&& \Phi_L = (\phi_L^+,\phi_L^0) \sim (1,2,1,1/2;0), ~~~ 
\Phi_R = (\phi_R^+,\phi_R^0) \sim (1,1,2,1/2;1/2).
\end{eqnarray}
As a result, the Yukawa terms $\bar{\psi}_L \Phi \psi_R$, $\bar{\psi}_L 
\tilde{\Phi}_L \nu_R$, $\bar{\psi}_R \tilde{\Phi}_R n_L$, $\bar{Q}_L 
\tilde{\Phi} Q_R$, $\bar{Q}_L \Phi_L d_R$, $\bar{Q}_R \Phi_R h_L$ are allowed, 
whereas $\bar{\psi}_L \tilde{\Phi} \psi_R$, $\bar{Q}_L \Phi Q_R$ are forbidden 
together with the bilinear terms $\bar{n}_L \nu_R$, $\bar{h}_L d_R$.
The breaking of $SU(2)_R \times U(1)_X \to U(1)_Y$ leaves $L = S + T_{3R}$ 
unbroken, so that $v_2 = \langle \phi_2^0 \rangle \neq 0$ [$\phi_2^0$ 
has $L=0$], but $\langle \phi_1^0 \rangle = 0$ [$\phi_1^0$ has $L=-1$].  
The former contributes to $m_e$ and $m_u$, whereas the latter means that 
$\nu_L$ and $n_R$ are not Dirac mass partners and can be completely 
different particles.  In fact, $m_\nu$, $m_d$ come from $v_3 = \langle \phi_L^0 
\rangle$, and $m_n$, $m_h$ from $v_4 = \langle \phi_R^0 \rangle$. This 
structure guarantees the absence of tree-level flavor-changing neutral 
currents.  As for the gauge bosons and their interactions, let 
$e/g_L = s_L = \sin \theta_W$, $e/g_R = s_R$, $e/g_X = \sqrt{1-s_L^2-s_R^2} 
= \sqrt{c_L^2 - s_R^2}$,  then
\begin{eqnarray}
&& A = s_L W_L^0 + s_R W_R^0 + \sqrt{c_L^2-s_R^2} X, \\ 
&& Z = c_L W_L^0 - (s_L s_R/c_L) W_R^0 - (s_L \sqrt{c_L^2-s_R^2}/c_L) X, \\ 
&& Z' = (\sqrt{c_L^2-s_R^2}/c_L) W_R^0 - (s_R/c_L) X, \\
&& g_Z = e/s_L c_L, ~~~ J_Z = J_{3L} - s_L^2 J_{em}, \\ 
&& g_{Z'} = e/s_R c_L \sqrt{c_L^2-s_R^2}, ~~~ J_{Z'} = s_R^2 J_{3L} + c_L^2 J_{3R} 
- s_R^2 J_{em}.
\end{eqnarray}
To avoid $Z-Z'$ mixing at tree level, the condition $v_2^2/(v_2^2+v_3^2) = 
s_R^2/c_L^2$ must be imposed.  In that case, $M_{W_R} \simeq (\sqrt{c_L^2-s_R^2} 
/c_L) M_{Z'}$.  Note that $W_R$ does not mix with $W_L$ because they have 
different $R$ parity.  In Fig.~1, the present Tevatron bound on $M_{Z'}$ 
is shown for various values of $s_R^2$, showing a typical bound of about 1 TeV.
\vspace{0.1cm}
\begin{figure}[htb]
\includegraphics[width=0.8\textwidth]{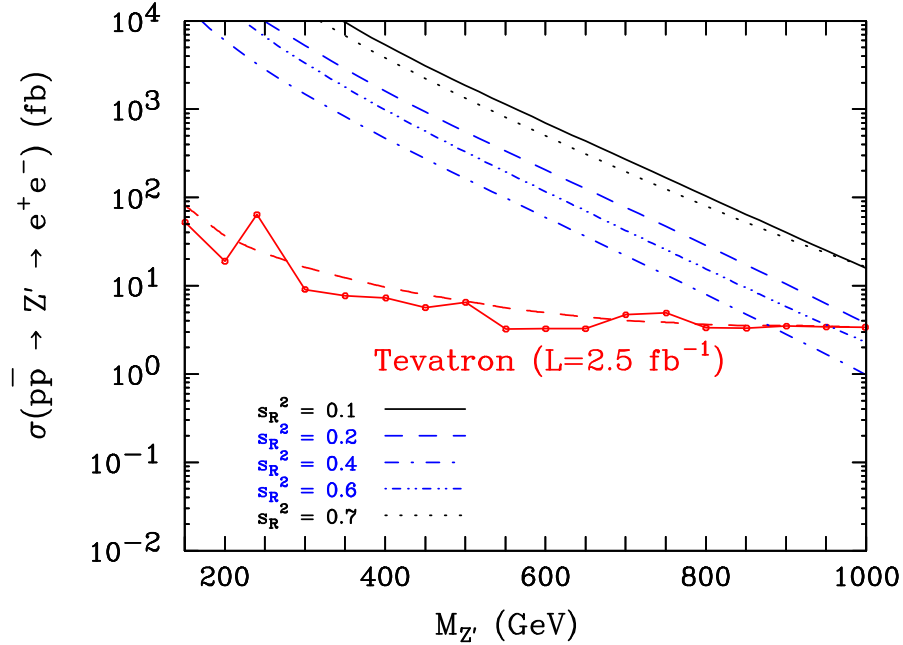}
\caption{$\sigma(p \bar{p} \to Z' \to e^+e^-)$ vs $M_{Z'}$  
compared against Tevatron data.} 
\end{figure}

\section{CDMS and More}
The usual leptons have $L=1$ as expected, but there are now new particles 
also with lepton number: $W_R^+,\phi_R^+,\phi_1^+$ have $L=1$ and $h$ has 
$L=-1$ as well as $B=1/3$.  Thus they all have odd $R$ parity, i.e. 
$R = (-)^{3B+L+2j} = -1$, even though the model is nonsupersymmetric.  
The scotino $n$ has $L=2$ and thus also odd $R$.  The lightest $n$ 
is a dark-matter candidate, and will be considered below in the context of 
recent data from the CDMS-II collaboration~\cite{cdms10}.  Two possible 
dark-matter signal events were observed with an expected background of 
$0.9 \pm 0.1$.  The most stringent bound on the elastic spin-independent 
scattering cross section of $n q \to n q$ occurs at $m_n = 70$ GeV, and 
it is $3.8 \times 10^{-8}$ pb.  In the DLRM2,
\begin{equation}
{\cal L} = {g_{Z'}^2 n_V \over M_{Z'}^2} (\bar{n} \gamma_\mu n)(u_V \bar{u} 
\gamma^\mu u + d_V \bar{d} \gamma^\mu d),
\end{equation}
where $n_V = c_L^2/4$, $u_V = c_L^2/4 - 5 s_R^2/12$, $d_V = s_R^2/12$.  Let 
$f_P = g_{Z'}^2 n_V (2 u_V + d_V)/M_{Z'}^2$, $f_N = g_{Z'}^2 n_V (u_V + 2 
d_V)/M_{Z'}^2$, then
\begin{equation}
\sigma_0 \simeq {4 m_P^2 \over \pi} {[Z f_P + (A-Z)f_N]^2 \over A^2}.
\end{equation}
Using $^{73}$Ge, i.e. $Z=32$ and $A-Z=41$, as an estimate, the CDMS bound 
(of $3.8 \times 10^{-8}$ pb at $m_n = 70$ GeV) implies a bound on $M_{Z'}$.  
In Fig.~2, the resulting lower bounds on $M_{Z'}$ from the Tevatron 
search and from CDMS are plotted as functions of $s_R^2$.\vspace{0.1cm}
\begin{figure}[htb]
\includegraphics[width=0.8\textwidth]{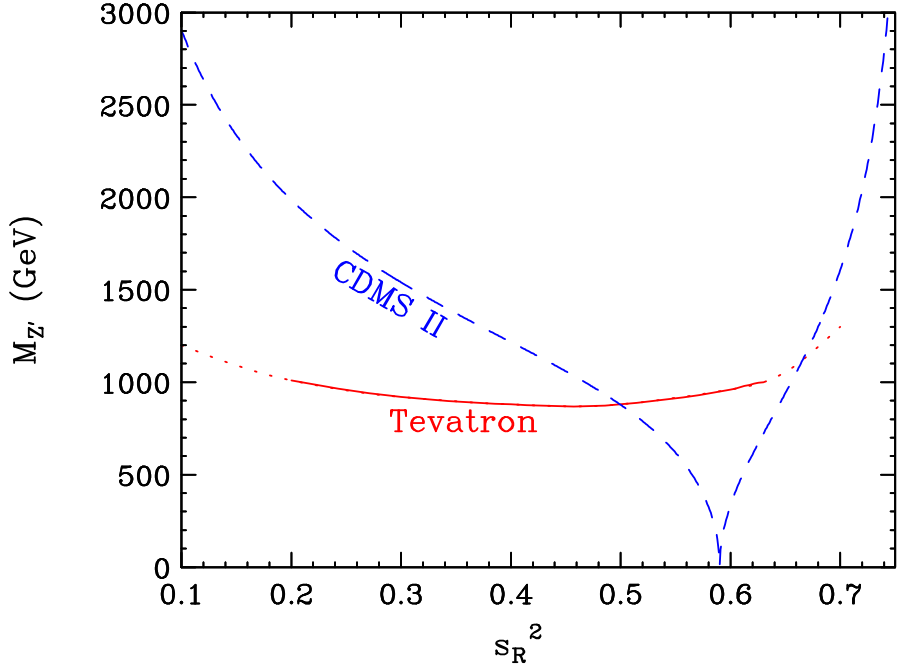}
\caption{Lower bounds on $M_{Z'}$ 
vs $s_R^2$ from the Tevatron search (red solid line) and from the CDMS  
search at $m_n = 70$ GeV (blue dashed line).  The dotted segments assume a 
simple extrapolation of the Tevatron data.} 
\end{figure}

To obtain the correct dark-matter relic abundance, the annihilation of 
$n \bar{n} \to Z' \to$ SM fermions is considered.  The thermally averaged 
cross section multiplied by the relative velocity of the annihilating 
particles is given by
\begin{equation}
\langle \sigma v_{rel} \rangle_{Z'} = {\pi \alpha^2 (3 - 9r + 10r^2) m_n^2 
\over 2 c_L^4 r^2 (1-r)^2 (4m_n^2 - M_{Z'}^2)^2},
\end{equation}
where $r = s_R^2/c_L^2$.  Fixing the above at 1 pb, the values of $m_n$ and 
$M_{Z'}$ are constrained as a function of $s_R^2$.  For $m_n = 70$ GeV, 
there is no solution, but if $m_n$ is greater than about 300 GeV, 
solutions exist which are consistent with the Tevatron bound as well as 
the CDMS bound.  In the range $0.3 < m_n < 1.0$ TeV, the latter is well 
approximated by $\sigma_0 < 2.2 \times 10^{-7}$ pb ($m_n$/1 TeV)$^{0.86}$. 
In Fig.~3, the cases $m_n=400$ and 600 GeV are shown.
\vspace{0.1cm}
\begin{figure}[htb]
\includegraphics[width=0.45\textwidth]{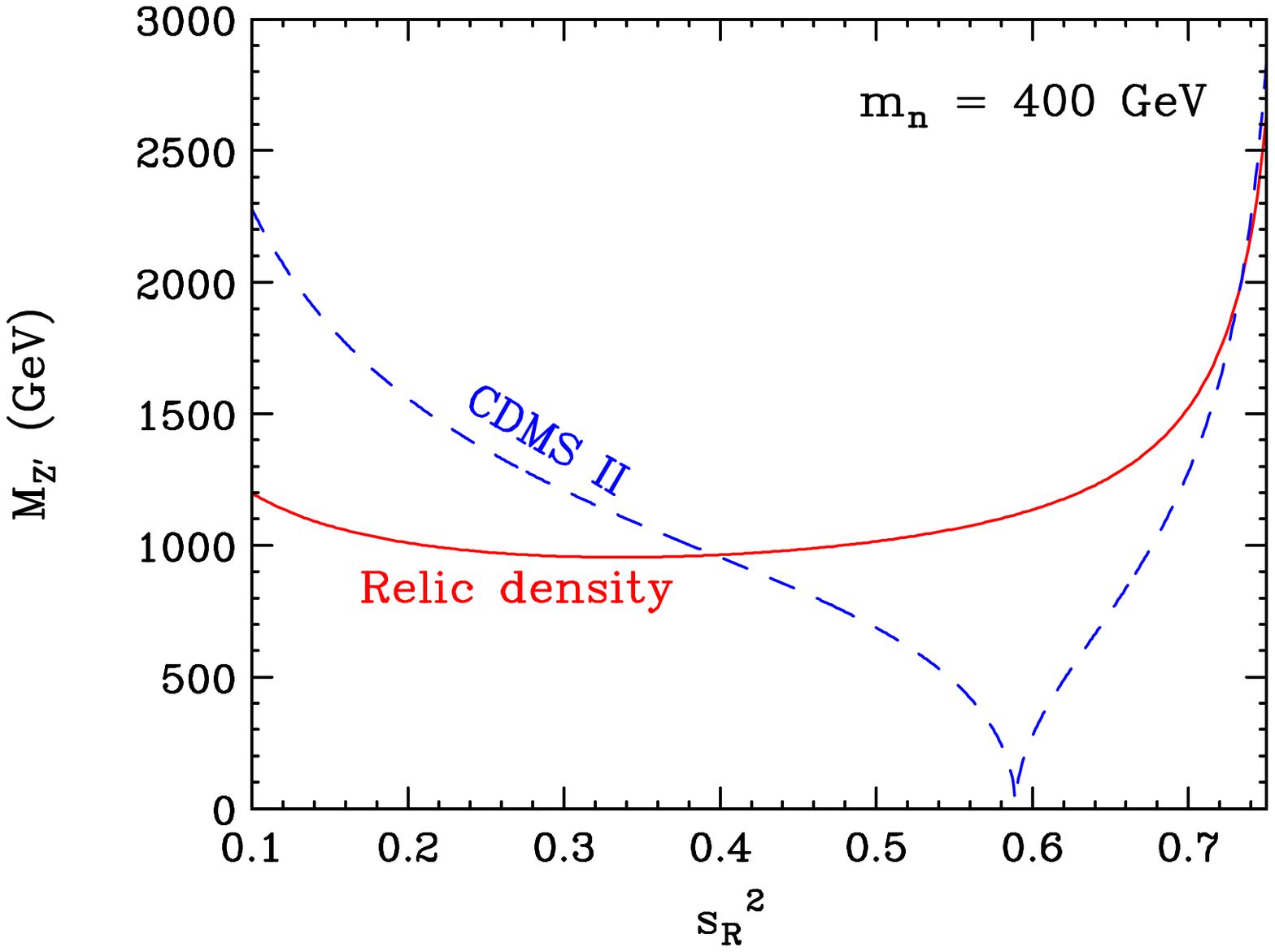} ~~~~~~
\includegraphics[width=0.45\textwidth]{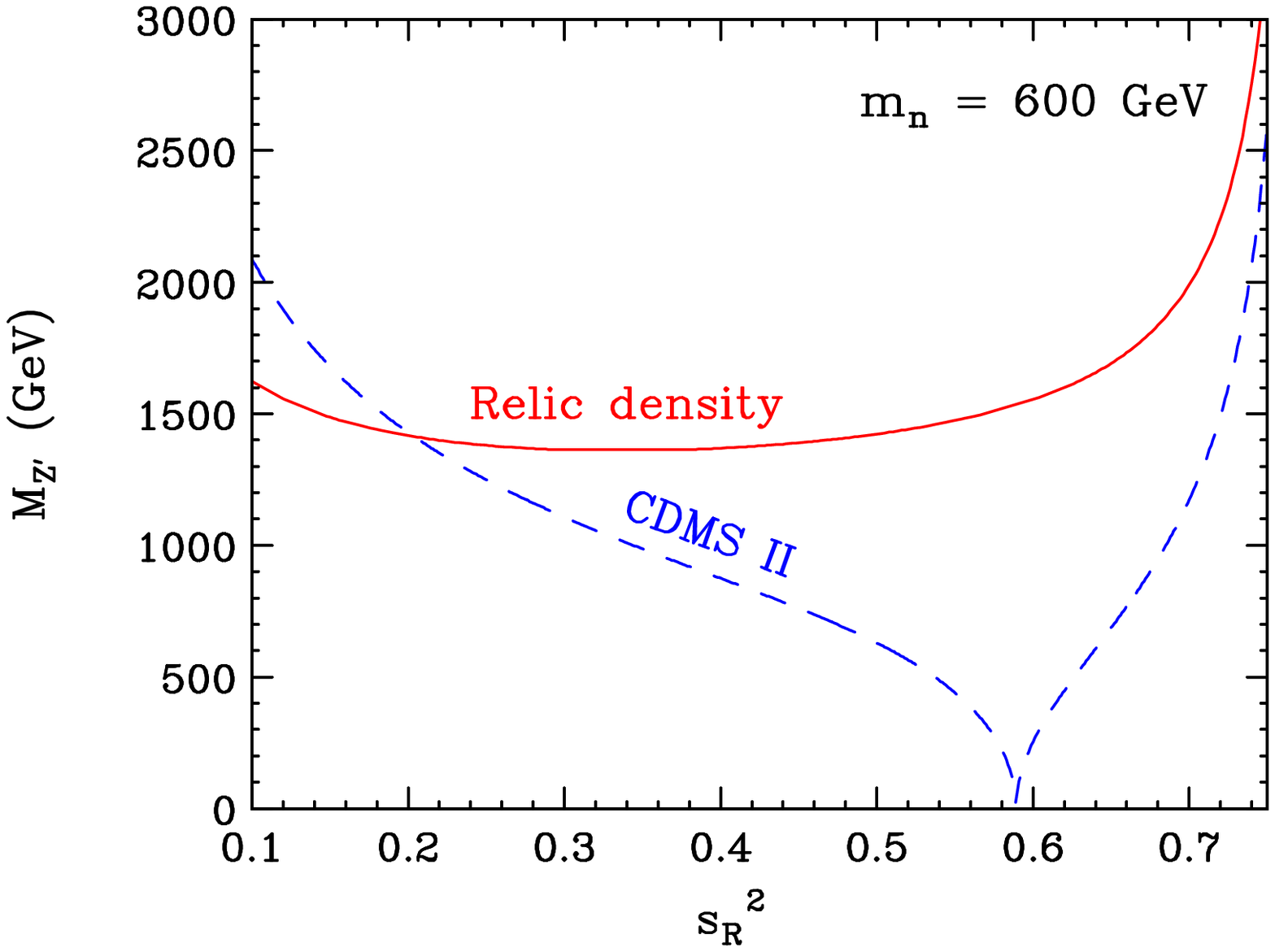} 
\caption{The CDMS bound on $M_{Z'}$ (blue dashed line) and its value 
(red solid line) from $\langle \sigma v_{rel} \rangle_{Z'} = 1$ pb vs 
$s_R^2$ for $m_n = 400$ and 600 GeV.} 
\end{figure}

The $n \bar{n}$ annihilation to leptons through $W_R^\pm$ exchange also 
contributes, i.e.
\begin{equation}
\langle \sigma v_{rel} \rangle_{W_R} = {3 g_R^4 m_n^2 \over 64 \pi (m_n^2 
+ M_{W_R}^2)^2},
\end{equation}
but it is subdominant and has been neglected.

\section{LHC and More}
At the LHC ($E_{cm} = 14$ GeV), $Z'$ may be discovered with 10 dilepton 
events.  Using the cuts
\begin{itemize}
\item{$p_T > 20$ GeV for each lepton,}
\item{ $|\eta| < 2.4$ for each lepton,}
\item{$|M_{l^-l^+} - M_{Z'}| < 3 \Gamma_{Z'}$,}
\end{itemize}
the SM background is negligible.  With an integrated luminosity of 1 
fb$^{-1}$, the discovery reach of the $Z'$ of the DLRM2 is about 2 TeV, 
as shown in Fig.~4.

\begin{figure}[htb]
\includegraphics[width=0.8\textwidth]{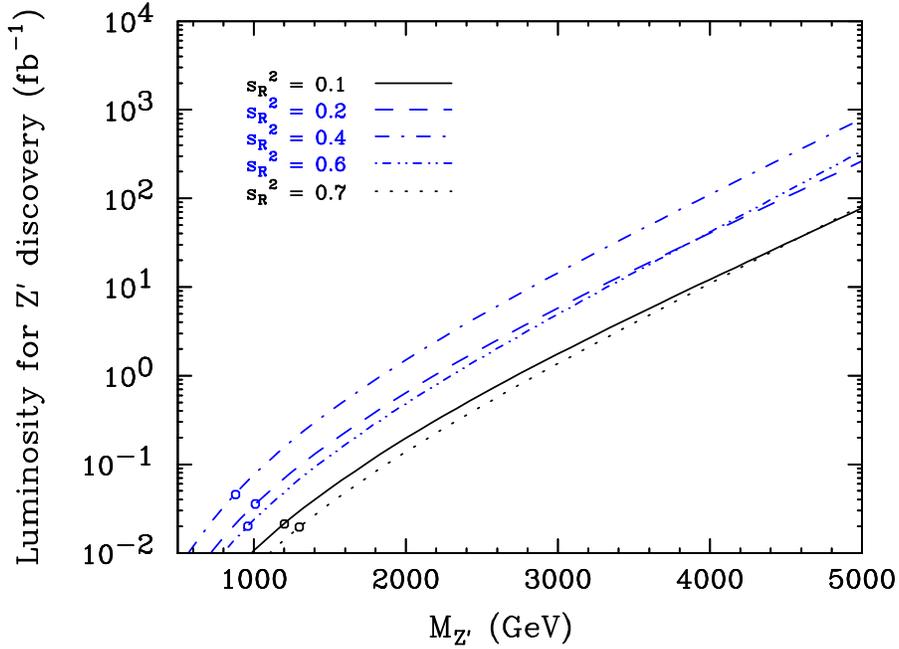}
\caption{Luminosity for $Z'$ discovery by 10 dielectron events at LHC. 
Small circles are Tevatron limits.}
\end{figure}

To distinguish this $Z'$ from others, the following ratios~\cite{gm08} 
may be considered:
\begin{eqnarray}
{\Gamma(Z' \to t \bar{t}) \over \Gamma(Z' \to \mu^- \mu^+)} &=& 
{(9 - 24r + 17r^2) \over 3(1 - 4r + 5r^2)} = 4.44~(g_L=g_R), \\ 
{\Gamma(Z' \to b \bar{b}) \over \Gamma(Z' \to \mu^- \mu^+)} &=& 
{5r^2 \over 3(1 - 4r + 5r^2)} = 0.60~(g_L=g_R),
\end{eqnarray}
where $r=s_R^2/c_L^2$.  In the conventional left-right model, the numerator 
for $b \bar{b}$ is changed to $(9 - 12r + 8r^2)$, i.e. 13.6 larger 
($g_L=g_R$).  In the ALRM, the denominator for both is changed to 
$3(2 - 6r + 5r^2)$, i.e. 2.6 larger ($g_L=g_R$).

There are also important loop effects~\cite{m00} on rare processes from the 
new interactions $\gamma W_R^+ W_R^-$, $Z W_R^+ W_R^-$, $W_R^+ \bar{n}_R e_R$, 
$W_R^+ \bar{u}_R h_R$, etc.  The anomalous magnetic moment of the muon 
receives a contribution of order $10^{-10}$, below the present experimental 
sensitivity of $10^{-9}$.  The flavor-changing radiative decay $\mu \to e 
\gamma$ has the branching fraction
\begin{equation}
B(\mu \to e \gamma) = {3 \alpha \over 32 \pi} \left( {s_L M_{W_L} \over 
s_R M_{W_R}} \right)^4 |\sum_i U_{\mu i} U_{e i} F_\gamma(r_i)|^2,
\end{equation}
where $r_i = m_{n_i}^2/M_{W_R}^2$ and 
\begin{equation}
F_\gamma (r_i) = {r_i (-1 + 5r_i + 2r_i^2) \over (1 - r_i)^3} + {6r_i^3 \ln r_i 
\over (1 - r_i)^4}.
\end{equation}
It is suppressed by $M_{W_R}$ and for $M_{W_R} = 1.17$ TeV (corresponding 
to $M_{Z'} = 1.4$ TeV), the current experimental bound of $1.2 \times 
10^{-11}$ implies $|\sum_i U_{\mu i} U_{e i} F_\gamma(r_i)| < 0.05$ for 
$g_L=g_R$.
A more sensitive probe of the existence of these new interactions is 
$\mu \to e e e$ or $\mu-e$ conversion in a nucleus or muonic atom~\cite{y10}.
The reason is that there is an effective $\mu \to e Z$ vertex from 
$Z W_R^+ W_R^-$ and $W_R^+ \bar{n}_R e_R$, given by
\begin{equation}
g_{\mu e Z} = {e^3 s_L \over 16 \pi^2 s_R^2 c_L} \sum_i U_{\mu i} U_{e i} F_Z(r_i),
\end{equation}
where $F_Z(r_i) = r_i/(1-r_i) + r_i^2 \ln r_i/(1-r_i)^2$, which is not 
suppressed if $r_i$ is not small, which holds even if the $SU(2)_R$ scale 
is much greater than the electroweak scale.  This unusual (nondecoupling) 
property depends crucially on the $Z W_R^+ W_R^-$ vertex, which is not 
available in other extensions of the SM, including all $U(1)'$ models. 
The current experimental bound of $1.0 \times 10^{-12}$ on 
$B(\mu \to e e e)$ implies $|\sum_i U_{\mu i} U_{e i} F_Z(r_i)| < 1.44 
\times 10^{-3}$ for $g_L=g_R$.

\section{Conclusion}
The presence of $\nu_R$ is unavoidable in a left-right gauge extension of the 
Standard Model.  However, it does not have to be the Dirac mass partner of 
$\nu_L$.  In that case, it should be renamed $n_R$ and could function as a 
scotino, i.e. a dark-matter fermion.  In the context of the recently proposed 
new dark left-right model (DLRM2), latest CDMS observations are shown to 
be consistent with the lightest $n$ at about a few hundred GeV in mass with  
the new $Z'$ gauge boson at less than 2 TeV.  The latter should then 
be accessible directly at the LHC, while the $W_R^\pm$ gauge bosn may 
contribute indirectly to enhancing rare lepton-flavor-changing processes 
such as $\mu \to e e e$ and $\mu-e$ conversion in a nucleus or muonic atom.

\section*{Acknowledgments} 
This work was supported in part by the U.~S.~Department of Energy under 
Grant No.~DE-FG03-94ER40837.  I thank D. Delepine and the other organizers 
of the VI International Workshop on the Dark Side of the Universe for 
their great hospitality and a stimulating meeting in Leon.





\bibliographystyle{aipproc}   


\end{document}